\documentclass[journal=langd5,manuscript=article]{achemso}

\usepackage{bm}

\usepackage{graphicx}  

\usepackage[version=3]{mhchem}

\author{B.T.~Huang}
\author{M.~Roger}
\author{M.~Bonetti}
\affiliation{Service de Physique de l'Etat Condens\'e, CEA-IRAMIS-SPEC, 
CNRS, UMR 3680, CEA Saclay, F-91191 Gif-sur-Yvette Cedex, France}

\author{T.J.~Salez}
\affiliation{Service de Physique de l'Etat Condens\'e, CEA-IRAMIS-SPEC, 
CNRS, UMR 3680, CEA Saclay, F-91191 Gif-sur-Yvette Cedex, France}
\altaffiliation{\'Ecole des Ponts ParisTech, 6 et 8 avenue Blaise Pascal, 
Champs-sur-Marne F-77455 Marne-la-Vall\'ee, France}

\author{C.~Wiertel-Gasquet}
\affiliation{Service de Physique de l'Etat Condens\'e, CEA-IRAMIS-SPEC, 
CNRS, UMR 3680, CEA Saclay, F-91191 Gif-sur-Yvette Cedex, France}

\author{E.~Dubois}%
\affiliation{Laboratoire PHysicochimie des Electrolytes et Nanosyst\`emes
 InterfaciauX (PHENIX), Sorbonne Universit\'es,
 UPMC Univ. Paris 06, CNRS, UMR 8234, F-75005, Paris France}

\author{R.~Cabreira Gomes}
\affiliation{Grupo de Fluidos Complexos, Instituto de Fisica \& 
Instituto de Quimica, Universidade de Bras\'{i}lia, 
CP 04478, 70904-970 Bras\'{i}lia (DF), Brazil}
\altaffiliation{Laboratoire PHysicochimie des Electrolytes et Nanosyst\`emes
 InterfaciauX (PHENIX), Sorbonne Universit\'es,
 UPMC Univ. Paris 06, CNRS, UMR 8234, F-75005, Paris France}

\author{G.~Demouchy}
\author{G.~M\' eriguet}
\author{V.~Peyre}
\author{M.~Kouyat\'e}
\affiliation{Laboratoire PHysicochimie des Electrolytes et Nanosyst\`emes
 InterfaciauX (PHENIX), Sorbonne Universit\'es,
 UPMC Univ. Paris 06, CNRS, UMR 8234, F-75005, Paris France}

\author{C.L.~Filomeno}
\affiliation{Grupo de Fluidos Complexos, Instituto de Fisica \& 
Instituto de Quimica, Universidade de Bras\'{i}lia, 
CP 04478, 70904-970 Bras\'{i}lia (DF), Brazil}
\altaffiliation{Laboratoire PHysicochimie des Electrolytes et Nanosyst\`emes
 InterfaciauX (PHENIX), Sorbonne Universit\'es,
 UPMC Univ. Paris 06, CNRS, UMR 8234, F-75005, Paris France}

\author{J.~Depeyrot}
\author{F.A.~Tourinho}
\affiliation{Grupo de Fluidos Complexos, Instituto de Fisica \& 
Instituto de Quimica, Universidade de Bras\'{i}lia, 
CP 04478, 70904-970 Bras\'{i}lia (DF), Brazil}

\author{R.~Perzynski}%
\affiliation{Laboratoire PHysicochimie des Electrolytes et Nanosyst\`emes
 InterfaciauX (PHENIX), Sorbonne Universit\'es, 
 UPMC Univ. Paris 06, CNRS, UMR 8234, F-75005, Paris France}
\author{S.~Nakamae}%
\email{sawako.nakamae@cea.fr}
\affiliation{Service de Physique de l'Etat Condens\'e, CEA-IRAMIS-SPEC,
CNRS, UMR 3680, CEA Saclay, F-91191 Gif-sur-Yvette Cedex, France}

\title{Experimental Study of Thermodiffusion and Thermoelectricity in Charged Colloids}

\begin{document}

\begin{abstract}
The Seebeck and Soret coefficients of ionically stabilized suspension of maghemite nanoparticles 
in dimethyl sulfoxide are experimentally studied as a 
function of nanoparticle volume fraction. 
In the presence of a temperature gradient, the charged colloidal nanoparticles 
experience both thermal drift due to their interactions with the solvent 
molecules and electric forces proportional to the internal thermoelectric 
field. The resulting thermodiffusion of nanoparticles is observed through 
Forced Rayleigh scattering, while the thermoelectric field is accessed through 
voltage measurements in a thermocell. Both techniques provide  
independent estimates of nanoparticle's entropy of transfer as high as 
75 meV.K$^{-1}$. Such a property may be used 
to improve the thermoelectric coefficients in liquid thermocells. 
\end{abstract}

\section{1. INTRODUCTION}
The thermoelectric effect (Seebeck effect) is known to influence the thermodiffusion behavior of charged colloidal suspensions with contributions coming from both electrolytes and charged colloidal particles themselves. \cite{Wurger2008108302, Majee2011061403, Sugioka} 
Under a thermal gradient $\bm{\nabla} T$, the thermal drift of ionic species, \emph{i}, induces concentration
gradients $\bm{\nabla}n_i/n_i= -\alpha_i\bm{\nabla}T$ (Soret effect) and an
internal electric field $\bm{E}=\mathcal{S}_e\bm{\nabla}T$ (Seebeck effect)
Both Soret ($\alpha_i$) and Seebeck ($\mathcal{S}_e$) coefficients depend 
on the \emph{Eastman entropies of transfer} $\widehat{S}_i$ 
\cite{Majee2011061403,Agar196331,Vigolo-Piazza,Sugioka,Putnam},
 which characterize the interaction of species $i$ with the solvent 
 \cite{Marcus2009}.  
The absolute value of $\widehat{S}_i$ generally 
increases with the ion size. 
Large Soret effects have indeed been reported experimentally in various 
colloidal 
suspensions such as silica particles, DNA molecules, polystyrene spheres and
 magnetic nanoparticles (ferrofluids) \cite{Ning200810927,Wiegand,
Duhr200619678,Stadelmaier20099147, Piazza2008153102, Völker20044315} 
reflecting the large Eastman entropy of transfer associated with their
equally large physical size (in the nm--$\mu$m range). The 
corresponding Seebeck coefficient, on the other hand, has rarely been 
studied \cite{Jennifer}.

Here, we investigate one such charged colloidal suspension, namely,
 ionically stabilized ferrofluids in dimethyl sulfoxide (DMSO). 
Ferrofluids were chosen not only because of the high Soret
 coefficients \cite{Völker20044315, Sprenger2013429} of nanoparticles (NPs)
but also because 
of their magnetic nature which may offer an additional control
parameter (magnetic field \cite{Blums2013}) on the thermodiffusive and 
thermoelectric behaviour. 
We determine independently the Eastman entropy of transfer of NPs
 $\widehat{S}$ through
 (i) the Soret effect using forced Rayleigh scattering measurements and 
(ii) the Seebeck effect using a thermocell (see experimental section below).
The values of  $\widehat{S}$ deduced from the two experiments
 agree quantitatively and are almost three orders of magnitude higher
 than that of typical ions in electrolytes \cite{Agar196331}.
Furthermore, we show that the thermodiffusion of NPs has a sizable influence on the Seebeck coefficient,
an effect that may be used in liquid thermoelectric applications \cite{Gunawan2013304}.

 \section{2. EXPERIMENTAL SECTION}
	\subsection{2.1. Samples}
	
We have used ferrofluids based on well-known maghemite
 $\gamma$-Fe$_2$O$_3$ nanoparticles (average diameter $d=6.7$~nm and 
log-normal size distribution 0.38, determined from room-temperature 
magnetization measurements) dispersed in DMSO.
 Nanoparticles are chemically synthesized first 
in water (see \cite{Massart} for methods). This dispersion gives a ferrofluid 
of positively charged NPs with nitrate counterions, which are then replaced 
by perchlorate ones following the method described in \cite{Ivan}. 
At the end of the process, DMSO is added instead of water to obtain 
electrostatically stabilized dispersion with positively charged NP surface. 
The concentration of free perchloric acid in the solution was kept constant 
at $\approx$12.5~mM. This value was determined from the 
conductivity measurement in the 
supernatant of DMSO ferrofluid obtained after ultracentrifugation 
(60000 rpm, 1h30) which separates the NPs from the solvent.

	\subsection{2.2. Thermodiffusion measurements}

The Forced Rayleigh scattering technique used to extract the Soret and the 
NPs diffusion coefficients is well described in Ref. \cite{Soret-exp}. 
 The heating light (Hg arc lamp, 100Hz modulation) creates the optical image 
of a grid in the sample. Owing to the optical absorption by the NPs, a
 temperature grating is induced in the sample. Then a NP concentration 
grating settles due to the Soret effect in a few seconds at the spatial scale
 of $\approx 50~\mu$m. Both gratings are detected by the diffraction of a weakly 
absorbing test laser beam. As the gratings of the temperature and the NPs 
concentration evolve on timescales differing by orders of magnitude, 
this technique enables the use of a \emph{``two-timescale''} model
\cite{Soret-exp}.
The Soret coefficient is deduced from the temporal 
modulation of scattered intensity at a constant spatial modulation of NPs 
concentration. The diffusion coefficient is determined by the relaxation time
 measurement of the concentration grating after the heating source is 
switched off.

\subsection{2.3. Thermoelectric measurements}

The Seebeck coefficient measurements were performed in a homemade thermocell consisting of a vertical, cylindrical Teflon cell (14~mm high and 6~mm diameter)
 with two ends sealed by sapphire windows, similar to the setup described in
Refs.  \cite{Zinovyeva2014426, Bonetti2011114513}.  The ferrocene/ferrocenium (F$_{C}$/F$_{C}^{+}$) redox couple (2/4~mM respectively) was added to the sample in order to permit the exchange of electrons between the electrodes and the ferrofluid.
 These chemicals were purchased from $Sigma Aldrich$;
 ferrocene (F$_{C}$, $98\%$) 
and ferrocenium-tetrafluoroborate (F$_{C}$BF$_{4}$, technical grade) and used as 
received. The sample preparation is performed in a glovebox under a
 nitrogen atmosphere. 
It should be noted that the co-existence
 of the redox couple and nanoparticles did not change the redox
 potential (\textit{cf.} Supporting Information I)
 or cause aggregation of NPs.  
 The experiments are carried out between 30~$^\circ$C and 50~$^\circ$C 
(mean temperature) with the temperature difference between the two electrodes 
$\Delta T_{elect} \approx 4.3~^\circ$C (10~$^\circ$C difference between the cell 
extremities). 
The open circuit voltage is:
 $\Delta V = - \Lambda\Delta T_{elect}$, where $\Lambda$ shall be referred to
as the \emph{``thermoelectric coefficient''}, to be distinguished from the 
Seebeck coefficient  $\mathcal{S}_e$.
The thermoelectric voltage was monitored over 2 hours between 
each temperature change. 
The diffusion coefficient $D$ (see inset in Figure \ref{courbes_ST}), gives 
the nanoparticle diffusion time $\tau=l^2/(4\pi D)$ of the order of two days 
for our thermocell. 
Thus the observed electromotive force and thermoelectric coefficient
 $\Lambda$ correspond 
to those of the \emph{initial state} \cite{Agar196331}.

\section{3. RESULTS AND DISCUSSION}
	\subsection{3.1. Theoretical considerations}
To analyse our experimental results, here we consider a colloidal solution containing a concentration \emph{n} 
of charged particles with a structural diameter $d$ and a structural 
charge $Z_{str}$ (here, particles are positively charged). 
They are stabilized in a monovalent electrolyte solution A$^+$B$^-$. 
Some anions B$^-$ are condensed within the
first solvation layers of the particles, partly canceling $Z_{str}$ 
thus leading to an effective charge $Z<<Z_{str}$ \cite{Belloni1997}.
The remaining anions-- whose Coulomb binding energy is smaller
than $k_BT$--are free. The concentrations of free anions and cations
 are $n_-$ and  $n_+$, respectively. The electroneutrality writes:
\begin{equation}
Zn+n_+-n_-=0.
\label{neutrality}
\end{equation}
Under a temperature gradient, the particle current $\bm{J_i}$
corresponding to the charged species $i$ is 
\cite{Wurger2008108302, Majee2011061403}:
\begin{equation}
\bm{J_i}=-D_i\left[\bm{\nabla} n_i + 
n_i\frac{\widehat{S}_i}{k_BT}\bm{\nabla} T -
n_i\frac{\xi_i e}{k_BT}\bm{E}\right],
\label{jcol}
\end{equation}
where $n_i$ is the particle density. The first term corresponds to  
Fick's diffusion with coefficient $D_i$, the second term, proportional to the
``Eastman entropy of transfer'' $\widehat{S}_i$,
\bibnote{{Here the ``Eastman entropy of transfer'' $\widehat{S}_i$ is understood from Onsager equations \cite{Agar196331}. $\widehat{S}_i=\overline{\overline{S}}_i-s_i$, where $s_i$ represents the partial molar entropy and $\overline{\overline{S}}_i$ is the coefficientof the particle current $\bm{J}_i$ in the expression of the entropy current $\bm{J}_S=\sum_i\overline{\overline{S}}_i \bm{J}_i-\kappa\bm{\nabla T}/T$; (the last term is the Fourier contribution with heat conductivity $\kappa$) 
}}
 represents the thermal drift,
and the last term is the electric drift in the presence of a local field
$\bm{E}$. The dimensionless number 
$\xi_i=k_BT{\mu^{el}_i}/{eD_i}$ is
proportional to the ratio of the electrophoretic mobility 
$\mu^{el}_i$ to the diffusion coefficient $D_i$ \cite{Majee2011061403}.
For small point-like ions
the Einstein relation is valid and $\xi_i$ is simply the ionic 
charge number $z_i$. 
The particle currents corresponding to small non-interacting monovalent
ions  are:
\begin{equation}
\bm{J}_\pm=-D_\pm\left[\bm{\nabla} n_\pm + 
n_\pm\frac{\widehat{S}_\pm}{k_BT}\bm{\nabla} T \mp
n_\pm\frac{e}{k_BT}\bm{E}\right].
\label{jion}
\end{equation}
For colloidal particles $\xi$ is of the same order of magnitude as--but
 not equal to--the effective charge $Z$ \cite{Majee2011061403}.
At large volume fractions,
the interaction between NPs needs to be considered. This can be described
in terms of isothermal osmotic compressibility \cite{Vigolo-Piazza}
 $\chi (\phi)$, $\phi=V_{np} n$ where $V_{np}=\pi d^3/6$ is the nanoparticle
 volume (\emph{cf.} Supporting information II).
The $\phi$ dependence of the parameters $\widehat{S}$ and $\xi$ in 
eq~\ref{jcol} appears as:
\begin{equation}			
\widehat{S} = \widehat{S}_{0} \chi ( \phi)
\qquad \textrm{and} \qquad
\xi = \xi_{0} \chi ( \phi).
\label{correlation_empirique_S_chapeau}
\end{equation}
Choosing a hard sphere model with Carnahan-Starling equation of 
state \cite{Carnahan1969635}, $\chi ( \phi)$ becomes:
\begin{equation}
\chi \left(\phi_{eff}\right) = 
\frac{\left( 1-\phi_{eff} \right)^{4}}
{1+4 \phi_{eff}+4 \phi_{eff}^{2}-4 \phi_{eff}^{3}+4 \phi_{eff}^{4}},
\label{def_chi}
\end{equation}
where $\phi_{eff}=\phi (d_{HS}/d)^3$ represents an effective volume fraction
corresponding to hard-sphere diameter $d_{HS}=d+2\lambda_D$, where
 $\lambda_D$ is the Debye length. 

When the stationary state ($^{st}$) is reached, 
each of the three currents expressed
in eqs~\ref{jcol} and \ref{jion} vanishes. Combining these
equations with the electroneutrality condition,  eq~\ref{neutrality},
 we obtain \cite{Wurger2008108302,Majee2011061403}:
\begin{equation}
\bm{E}^{st} = \frac{1}{e} \left[ \frac{Zn \widehat{S}+
n_{+}\widehat{S}_{+}-n_{-}\widehat{S}_{-}}{n_{+}+n_{-}+\xi Zn} \right]
 \bm{\nabla} T = \mathcal{S}^{st}_{e} \bm{\nabla} T.
\label{E_st}
\end{equation}
Substituting this expression in the nanoparticle current 
(eq~\ref{jcol}), we obtain:
\begin{equation}
\frac{\bm{\nabla} n}{n}=-\frac{1}{k_BT}
(\widehat{S}-\xi e \mathcal{S}^{st}_{e})\bm{\nabla} T=-\alpha \bm{\nabla T},
\end{equation}
where 
\begin{equation}
\alpha=(\widehat{S}-\xi e \mathcal{S}^{st}_{e})/k_BT 
\label{alphaSoret}
\end{equation}
is the Soret coefficient.
For uncharged particles only the first term is present.
In a series of recent papers, W\" urger and coauthors 
\cite{Wurger2008108302,Majee2011061403,WurgerLangmuir,Fayolle2008041404}
 have emphasized the
importance of the second term in charged colloidal suspensions.
 Recent experiments
 by Eslahian \textit{et al.}  \cite{Eslahian2014} 
 on the salinity (electrolytes) effects on the
 thermodiffusion of polystyrene sulfonate beads appear to confirm these
 theoretical claims.

	\subsection{3.2. Thermodiffusion and thermoelectric analysis}

\begin{figure}[!b]
\begin{center}
\includegraphics[scale=0.35]{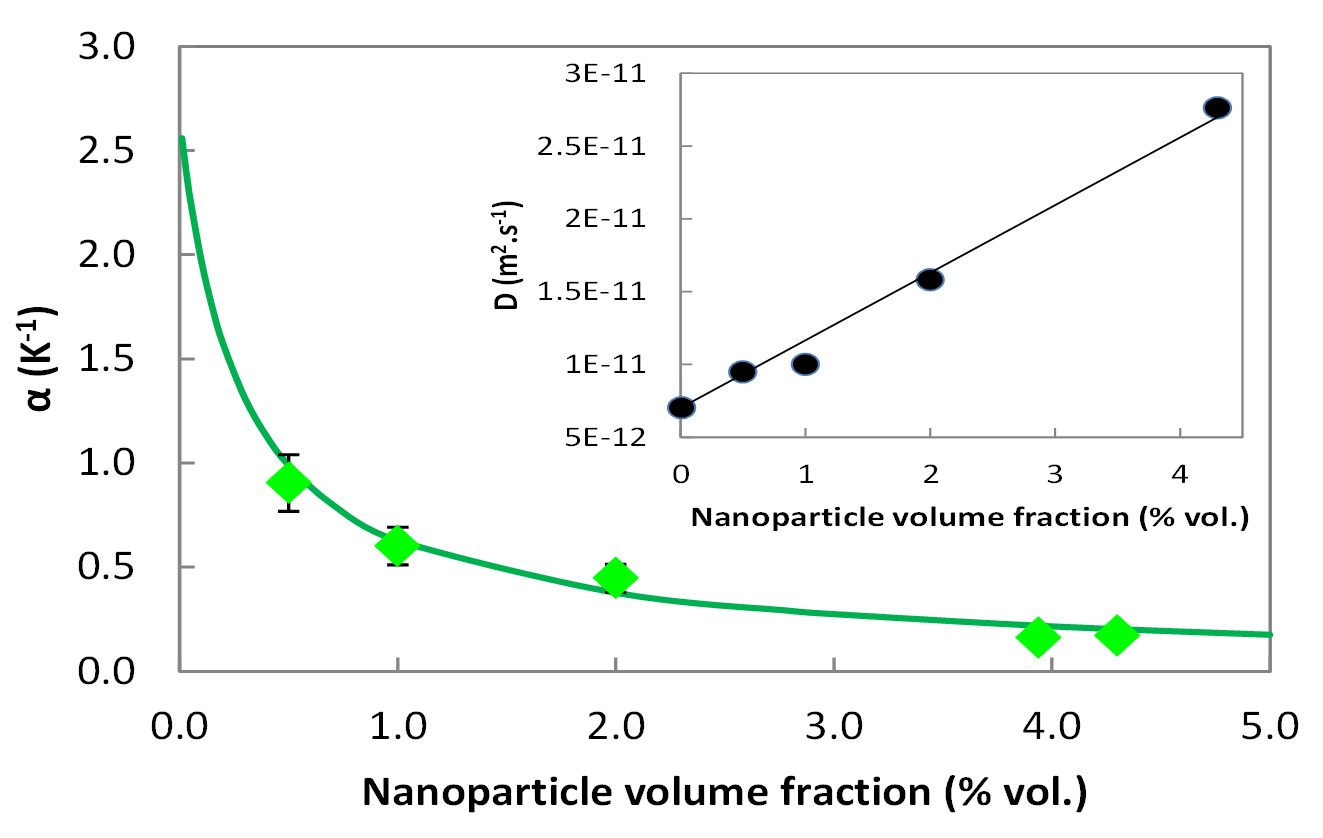}
\caption{Soret coefficient $\alpha$, and diffusion coefficient $D$ (inset)
 as a function of NP volume 
fraction. Note that the measurements at higher concentrations ($\approx 4\%$) 
 were performed on another set of DMSO based ferrofluids  in 
similar ionic conditions. }
\label{courbes_ST}
\end{center}	
\end{figure}

The Soret coefficient $\alpha$, determined at different NP volume
fractions $\phi$ are 
reported in Figure~\ref{courbes_ST} with an inset showing linear variation of the
diffusion coefficient:
$
D(\phi) \approx 7.03\times 10^{-12}(1 + 66.02\phi)~\rm{m}^2.\rm{s}^{-1}.
$
The variation in $\alpha$ was analyzed using eqs~\ref{E_st}
 and \ref{alphaSoret}, with the approximation (to be justified later):
 $\widehat{S}  >> \xi_0\widehat{S}_{+}, \xi_0\widehat{S}_{-}$ \textit{i.e.}:
\begin{equation} 
\alpha \approx \frac{1}{k_{B}T} \left[ 
\frac{  \widehat{S}(\phi_{eff})  \left( 1+ {Z \tilde{n}}/{2} \right) }
{1+(\xi (\phi_{eff})+1)Z\tilde{n}/2} \right].		
\label{expression_ST_simple}
\end{equation}
Here, we define $\tilde n=n/n^+$, where $n^+$ (H$^+$ ions) is kept constant.
The nanoparticle structural diameter is $d=6.7~$nm, the Debye
length is $\lambda_D = 2.1~nm$ in a solution of 12.5~mM HClO$_4$ in DMSO, 
with dielectic constant $\epsilon=48$, at room temperature.
$\xi_0 \approx 25$ was estimated from the measurement of the electrophoretic
mobility of a NP suspension at $\phi=0.05$\% using the laser Doppler 
velocimetry technique (NanoZS Malvern GB). 
The remaining unknown parameters $Z$ and  $\widehat{S}_0$ are determined through
the fit  (solid line in Figure \ref{courbes_ST}) 
of the experimental data by eq~\ref{expression_ST_simple}.
We obtain: $Z\approx 30$ and $\widehat{S}_0\approx 68~$meV.K$^{-1}$,
or equivalently $\widehat{S}_{0}/\xi_0 \approx 2.7$~meV.K$^{-1}$,
 one order of magnitude 
higher than typical values corresponding to electrolytes \cite{Agar196331},
which justifies our previous approximation:
 $\widehat{S}  >> \xi_0\widehat{S}_{+}, \xi_0\widehat{S}_{-}$.

\begin{figure}[!b]
\begin{center}
\includegraphics[scale=0.34]{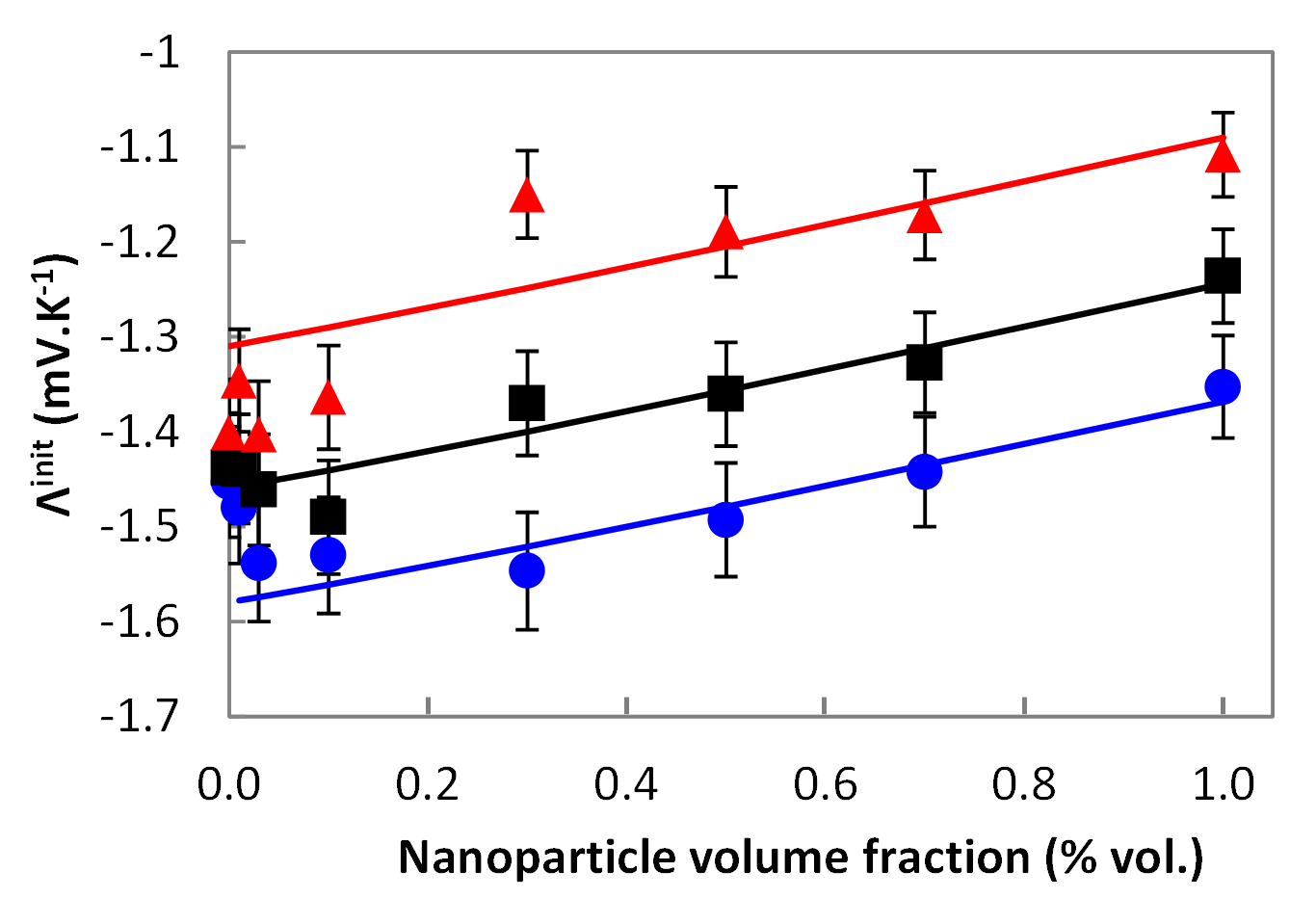}
\caption{ $\Lambda^{init}$ measured as a function of NP volume 
fraction $\phi$ at cell median temperature of 30 (circles), 40 (squares) 
and 50~$^\circ$C (triangles). The 
solid lines are fits to the Eq.~(\ref{fitS}). See text for more 
explanation.}
\label{courbes_Seebeck}
\end{center}	
\end{figure}


In thermoelectric measurements, a temperature gradient 
$\bm{\nabla} T$ is established across a previously 
 isothermal, homogenous electrolyte. 
While the bulk distribution of different species is still uniform, 
an internal electric field, $\bm{ E}^{init}$ settles immediately within the 
fluid, resulting from charge accumulations at the cell boundaries 
\cite{Agar196331}.
This is due to the response of ions to thermal forces 
$\bm{f}_i=\widehat{S}_i\bm{\nabla}T$.
In open circuit conditions the total electric current:
$\bm{J}_{elec}= Ze\bm{J}+e\bm{J}_+-e\bm{J}_-$ is zero. 
Substituting eq~\ref{jcol} and eq~\ref{jion} for the particle currents
and taking into account the initial condition 
$\bm{\nabla}n_i=0$ we obtain \cite{Agar196331}:
\begin{equation}
\bm{ E}^{init} = 
\sum_{i}\frac{t_{i} \widehat{S}_{i}}{\xi_{i}e}
\bm{\nabla} T = \mathcal{S}_e^{init}\bm{\nabla} T,
\label{E_initial}
\end{equation}
where $t_i=\sigma_i/\sigma_T$ is the Hittorf transport number
of ionic species $i$, \textit{i.e} the relative contributions of its
conductivity $\sigma_i$ to the total conductivity: 
$\sigma_T=\sum_i\sigma_{i}$.

In a thermocell, where a reversible redox reaction occurs at the
 electrodes,  the difference of electrochemical potential 
between the hot and the cold electrodes at initial state is:
$\Delta \tilde{\mu}=-e \Delta V^{init}=e\Lambda^{init} \Delta T$, with
 \cite{Agar196331}  (see Supporting Information III)
 for more detail):
\begin{equation}
\Lambda^{init} = 
\frac{\Delta s_r}{e}+\mathcal{S}_e^{init}=
\frac{\Delta s_r}{e} + 
 \sum_{i} \frac{t_{i} \widehat{S}_{i}}{\xi_{i}e}.
\label{Se_initial}
\end{equation}
The first term $\Delta s_r=s_{Fc^+}-s_{Fc}$ represents the redox reaction entropy
at electrodes (\textit{i.e.} the difference of the partial molar entropies of 
F$_C^+$ and F$_C$) which remains constant through out the measurements. The second term arises
from the initial electric field as described in eq~\ref{E_initial}.

The measured $\Lambda^{init}$, as a function of nanoparticle volume fraction 
($\phi$), at different temperatures, are shown in Figure \ref{courbes_Seebeck}. 
The measurements are repeated at least 5 times at each 
concentration and temperature. The data dispersion is less than 4$\%$. The
$\Lambda^{init}$ values were found negative, as it can be expected from the 
negative
 redox reaction entropy of the F$_C$/F$_C^{+}$ couple \cite{Weaver1984},
 and varied  between -1.1 and -1.6 mV.K$^{-1}$. 
The absolute value of $\Lambda^{init}$
decreases with increasing $\phi$ as well as with the mean
cell temperature. 

\begin{figure}[!t]
\begin{center}
\includegraphics[scale=0.37]{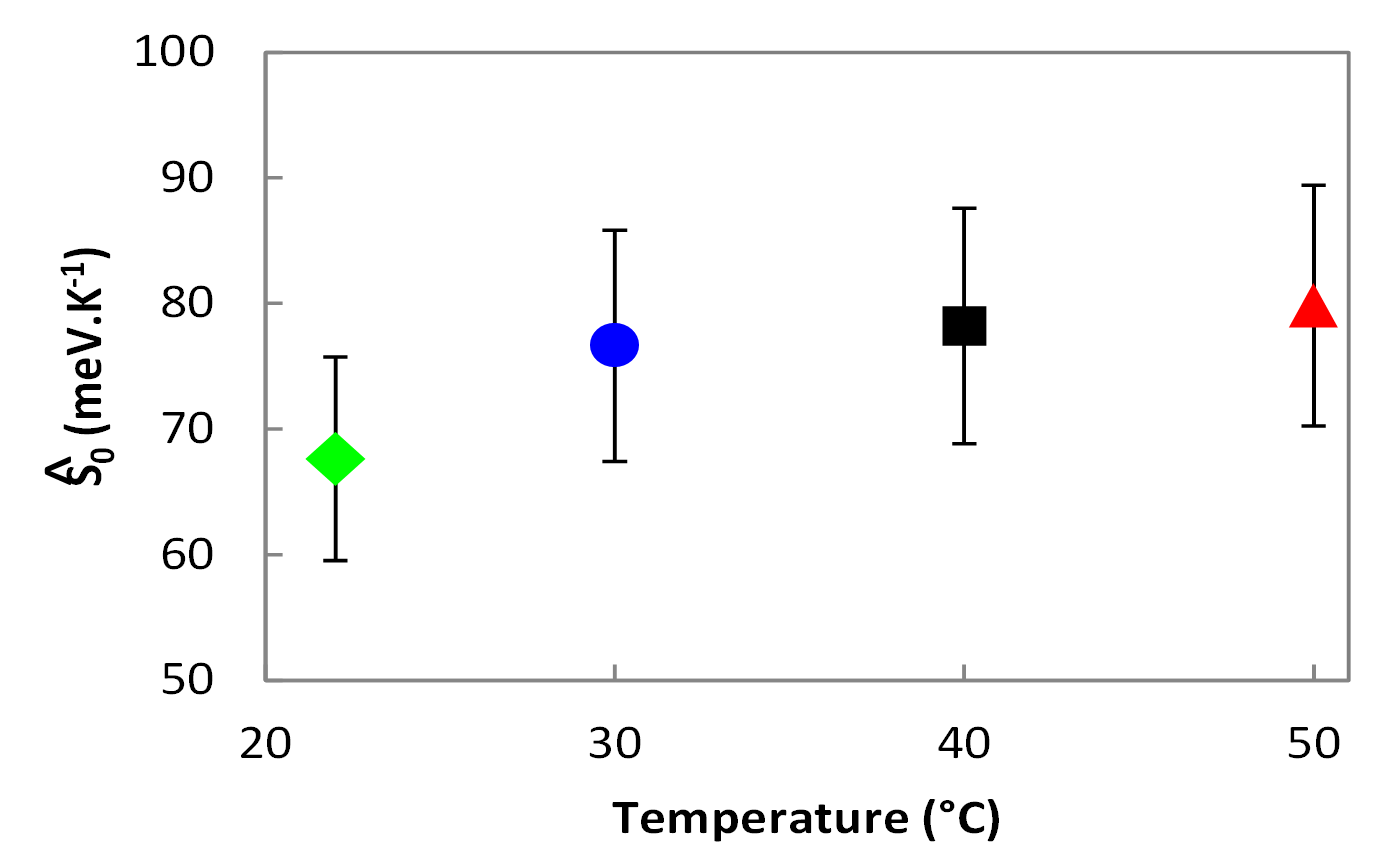}
\caption{Eastman transport entropy per NP in infinite dilution 
limit $\widehat{S}_{0}$ as a function of temperature, 
estimated by Seebeck
 (circle, square and triangle) and Soret (diamond) coefficients based models. 
The large error bars are mainly due to the uncertainty in the average
NP size (see text).
}
\label{courbes_S_chapeau_0}
\end{center}	
\end{figure}

At a fixed mean cell temperature and constant HClO$_4$
and  F$_C$/F$_C^{+}$  concentrations, the variations in $\Lambda^{init}$ stems
 from the term
${t\widehat{S}(\phi)}/({\xi (\phi)}e)$ of nanoparticles.
The nanoparticles' contribution to the electrical conductivity is:
$
\sigma= \xi (\phi) Z e^2 n D(\phi)/k_BT
$
is less than a few percent of the total conductivity
$\sigma_T$ at $\phi \le 1\%$. The
$\phi$ dependence of $\sigma_T$   was thus neglected. Combined together, 
eq~\ref{Se_initial} can be rewritten as:
\begin{equation}
\Lambda^{init}(\phi)=\Lambda^{init}(0)
+\frac{Ze}{k_BT} \frac{\phi}{V_{np}}\frac{D(\phi)}
{\sigma_T}\widehat{S}(\phi_{eff}).
\label{fitS}
\end{equation}
At room temperature, $D(\phi)$ is taken from the linear relation observed 
through Forced Rayleigh scattering (Figure \ref{courbes_ST} inset).  
$\sigma_T=35.5$~mS.m$^{-1}$ was measured at  room temperature.
 Since the dominant temperature
dependence of both $D$ and $\sigma_T$ arises from a same quantity \textit{i.e.}:
 the inverse friction coefficient $1/\eta (T)$ of the solvent, we shall take,
as a first approximation, $D/\sigma_T$ independant of $T$.
The $\phi$ dependence of  $\widehat{S}(\phi)$ is obtained from the hard sphere
model (eq~\ref{correlation_empirique_S_chapeau}). 
With $Z=30$ and $d=6.7~$nm, the experimental results (Figure \ref{courbes_Seebeck})
are fitted to eq~\ref{fitS} to deduce $\widehat{S}_{0}$. 
The results are compared in Figure \ref{courbes_S_chapeau_0} 
to the $\widehat{S}_{0}$ value determined
from the forced Rayleigh scattering experiments at 23$^o$C.
In the explored temperature range
  $\widehat{S}_{0}$ determined from the Soret
 and the Seebeck  coefficients measurements  
are  $\approx 75$~meV.K$^{-1}$, \textit{i.e.}
three orders of magnitude higher than the Eastman entropy of transfer of
small usual electrolyte ions (\textit{e.g.}, 0.12~meV.K$^{-1}$ for 
sodium ions  in water\cite{Agar196331}).
This observation supports the idea that the Eastman entropy of 
transfer play a major role in both thermoelectric and thermodiffusive 
phenomena in charged colloidal solutions.

\section{4. CONCLUSION}
In this work, we have measured the thermoelectric
 coefficient of an ionically stabilized 
ferrofluid as a function of nanoparticle volume fraction and compared the 
results to the corresponding Soret effect measurements. As expected, both
 coefficients depend on the concentration of charged nanoparticles and the
 values of Eastman entropy of transfer, $\widehat{S}_{0}$, determined from
 both experiments are found to be in fair quantitative agreement.
 Our results lend strong
 support to the existing theoretical models describing charged colloidal 
solutions' thermoelectric and thermodiffusive properties that both depend 
on $\widehat{S}_{0}$. 
Following the same rationale, one can postulate that the sign and the magnitude
 of Seebeck and Soret coefficients must depend on several experimental 
parameters: \textit{e.g.} the relative importance of the
 Eastman entropy of transfer betwen nanoparticles and surrounding 
ions,  the concentration of $all$ charged species in 
the solution, and the surface charge of colloidal particles. 
These extensive parameters can be tuned experimentally to control 
the thermoelectric coefficient $\Lambda$ of charged colloidal suspensions, offering a new perspective in future liquid thermocell research. 

\begin{acknowledgement}
This work was supported by ANR TEFLIC (Grant No. ANR-12-PRGE-0011-01), 
CEA program FLUTE (DSM-ENERGIE), LABEX-PALM (MAGTEP), 
CAPES-COFECUB n$^o$714/11 and PICS-CNRS  n$^o$5939.
The members of SPEC thank D. Duet and V. Padilla for technical assistance.
\end{acknowledgement}

\begin{suppinfo}
(I) Hard core interactions, (II) Cyclic voltammograms and 
(III)  Initial thermoelectric coefficient. 
\end{suppinfo}

\bibliography{biblio_botao}

\begin{mcitethebibliography}{30}
\bibitem[Not()]{Note-1}
{Here the ``Eastman entropy of transfer'' $\widehat{S}_i$ is understood from
  Onsager equations \cite{Agar196331}.
  $\widehat{S}_i=\overline{\overline{S}}_i-s_i$, where $s_i$ represents the
  partial molar entropy and $\overline{\overline{S}}_i$ is the coefficientof
  the particle current $\bm{J}_i$ in the expression of the entropy current
  $\bm{J}_S=\sum_i\overline{\overline{S}}_i \bm{J}_i-\kappa\bm{\nabla T}/T$;
  (the last term is the Fourier contribution with heat conductivity $\kappa$)
  }\relax
\mciteBstWouldAddEndPuncttrue
\mciteSetBstMidEndSepPunct{\mcitedefaultmidpunct}
{\mcitedefaultendpunct}{\mcitedefaultseppunct}\relax
\EndOfBibitem
\end{mcitethebibliography}

\end{document}